\documentstyle[preprint,tighten,aps,pra,amstex,psfig]{revtex}

\newcommand{\sinc}{\operatorname{sinc}}
\begin{document}
  
\draft

\author{Beno{\^\i}t Gr{\'e}maud  and Dominique Delande}
\address{Laboratoire Kastler Brossel, Universit{\'e} Pierre et
Marie Curie, T12, E1 \\
4, place Jussieu, 75252 Paris Cedex 05, France}
\title{Ghost orbits in the diamagnetic hydrogen spectrum
  using harmonic inversion}
\date{\today}
\maketitle
\begin{abstract}
The harmonic inversion method is applied in the case of the hydrogen atom in
a magnetic field to extract classical information from the
quantum photo-ionization cross-section. The study is made close to a
saddle-node bifurcation for which the usual semi-classical
formulas give diverging contributions. All quantities (actions,
stabilities and Maslov indices) for real orbits above the bifurcation
and for 
complex ghost orbits below the bifurcation, 
are found to be in excellent agreement with the
modified semi-classical predictions based on a normal form approach.

\end{abstract} 


\section{Introduction}

The hydrogen atom in a strong magnetic field is one of the most
appealing system to 
study chaotic effects in quantum mechanics: it has the minimum number 
of degrees of freedom, all quantum and classical properties can be
computed in a exact way~\cite{Friedrich89,Houches89}.  
Some other systems share these properties, 
but the difference is that the chaotic regime for the magnetized hydrogen 
atom is also observed experimentally~\cite{Holle88,Main94}, which is
the case of only few systems, as helium atom\cite{Domke95,Puttner99},
hydrogen atom in a microwave 
field\cite{Koch95,Bayfield91}, resonant tunneling
diode~\cite{Fromhold94,Muller95,Saraga98}, experiments with cold
atoms\cite{Moore94,Klappauf98}. Chaos being defined at the 
classical level as exponential sensitivity on initial conditions, a
major step in the 
understanding of chaos at the quantum level has been made trough the
Gutzwiller trace formula, linking the quantum density of states to the
classical periodic orbits of the system\cite{Gutzwiller90}. Later on,
similar formula were 
derived to explain experimental observation, like photo-ionization
cross-section, in terms of classical orbits (closed orbits in this
case)~\cite{Bogomolny89,Gao92}. 

The agreement between the numerical and experimental quantum results
was found to be good, except for parameters (energy, magnetic field)
too close to bifurcation points, at which the semi-classical
amplitudes are diverging. However, these divergences have been later
understood and classified using a normal form
approach~\cite{Main94,Mao92b}. Especially, it 
emphasized the contribution of complex ``ghost" \cite{Kus93}
orbits (where both position and
momentum are made complex quantities) in the photo-ionization
cross-section. This again was found to be in a qualitatively good
agreement with 
numerical and experimental data. Still, the comparison between the
quantum results and the classical predictions being obtained by
Fourier transforms of the quantum data, the accuracy is limited by the 
length of the available spectra (experimental or theoretical). For
example, the imaginary part of the action of the complex orbit being given
by the width of the corresponding peak in the Fourier transform, its
exact value is hidden by the broadening of this peak. For the same
reason, it is impossible to distinguish peaks closer one to each other 
than the ``Fourier limit'' as it is the case for the two orbits
created at a bifurcation point. 

In this paper, we show how an excellent agreement between the
quantum and the semi-classical properties in the vicinity of a
bifurcation can be 
obtained using the harmonic inversion, 
a newly developed method, which is able to bypass the Fourier
limitation~\cite{Neuhauser90,Wall95,Main97b}. The paper is divided as
follows~: in section~\ref{hydrogen} the 
essential properties of the hydrogen atom in a magnetic field are given,
then in section~\ref{bifu} we 
present the usual semi-classical approach of the saddle-node
bifurcation.  In section~\ref{harm} we briefly explain the harmonic inversion
method, which is applied in section~\ref{res} to follow, from our
quantum calculation, the properties of classical orbits through a
saddle-node bifurcation.

\section{Hydrogen atom in magnetic field}
\label{hydrogen}

In atomic units, the Hamiltonian of an hydrogen atom in a magnetic
field is given by (using cylindrical coordinates)~:

\begin{equation}
  H=\frac 12 {\mathbf{p}}^2-\frac 1r+\frac 12\gamma L_z+\frac 18\gamma^2\rho^2
\end{equation}
where $\gamma=B/B_0$, with $B_0=2.35\times10^5 T$. Due to the rotational
invariance around the $z-$axis of the Hamiltonian, $L_z$ is a good
quantum number and we shall take $L_z=0$ in what follows. 

The classical counterpart of this Hamiltonian has a scaling property,
that is, if we define new variables by~:
\begin{equation}
\label{scaling}
  \left\{\begin{array}{l}
\tilde{\mathbf{r}}=\gamma^{2/3}\mathbf{r} \\
\tilde{\mathbf{p}}=\gamma^{-1/3}\mathbf{p} \\
\tilde{t}=\gamma t \\
\end{array}\right.
\end{equation}
we obtain a new Hamiltonian $\tilde{H}$ given by~:
\begin{equation}
  \tilde{H}=\gamma^{-2/3}H=\frac {{\tilde{\mathbf{p}}}^2}2 
  -\frac 1{\tilde{r}}+\frac{\tilde{\rho}^2}8. 
\end{equation}
which does not depend anymore on $\gamma$. The classical dynamics of
this Hamiltonian is entirely fixed by the scaled energy $\epsilon$
given by~:
\begin{equation}
  \label{eet}
  \epsilon=\gamma^{-2/3}E.
\end{equation}
All properties of the classical trajectories of the original
Hamiltonian can be deduced from the scaled dynamics using the scaling
transformation~\eqref{scaling}. For example, the action $S$ of
an orbit (i.e. $\int \mathbf{p}.d\mathbf{q}$) is related to the reduced
action $\tilde{S}(\epsilon)$ (all quantities with $\tilde{}$ refer to
the scaled Hamiltonian) by~:
\begin{equation}
  S(E,\gamma)=\gamma^{-1/3}\tilde{S}(\epsilon).
\end{equation}

From the quantum point of view, this scaling introduces an effective
$\hbar$ value, which is easily seen on the scaled Schr{\"o}dinger
equation, $\tilde{H}\psi=\epsilon\psi$, for fixed scaled energy
$\epsilon$~: 
\begin{equation}
  \bigl(-\frac{\gamma^{2/3}}2\Delta_{\tilde{\mathbf{r}}}
  -\frac{1}{\tilde{r}}+
  \frac{\tilde{\rho}^2}8\bigr)\psi=\epsilon\psi.
\end{equation}
Thus, the effective $\hbar$ is given by $\gamma^{1/3}$ and so at
fixed value of the scaled energy $\epsilon$, the semi-classical limit
is obtained  when $\gamma$ tends to 0.

\section{Semi-classical approach and saddle-node bifurcation}
\label{bifu}

The semi-classical approximation of the
photo-ionization cross-section $\sigma(\gamma,E)$ is obtained by a
generalization 
of the Gutzwiller trace formula~\cite{Gutzwiller90,Gaspard95} and  has the
following expression~:
\begin{equation}
  \sigma(\gamma,E)=\sigma_{\mathrm{Coul}}(E)+\sigma_{\mathrm{osc}}(\gamma,E).
\end{equation}
$\sigma_{\mathrm{Coul}}(E)$ is a smooth background term and corresponds to the
Coulombic cross-section that would be obtained in the absence of a magnetic
field and, for energies close to the ionization threshold $E=0$,
$\sigma_{\mathrm{Coul}}(E)$ is almost independent from $E$, 
so that in first approximation
$\sigma_{\mathrm{Coul}}(E)\approx\sigma_{\mathrm{Coul}}(0)$. 
The second term is the oscillatory
part of the cross-section and is 
given as a sum other all orbits closing at the
nucleus (and their repetitions)~\cite{Bogomolny89,Gao92,Du88}~:
\begin{equation}
  \label{eq:trace}
  \sigma_{\mathrm{osc}}(\gamma,E)=\frac{4\pi\omega}{c}\sum_k A_k(\gamma,E)
  \sin\left(\phi_k(\gamma,E)\right),
\end{equation}
$\omega$ being the photon frequency. Actually, looking at
cross-section for energies 
close the ionization threshold (i.e. $E=0$), $\omega=(E-E_i)/\hbar$ is
almost given  
by the initial state energy $-E_i$ and does not depend on the energy $E$.
In the preceding formula, each amplitude $A_k$ involves different classical
quantities~\cite{Mao92a}, among which the matrix element $m_{12}$ of
the monodromy matrix (i.e. the stability matrix restricted to
deviations perpendicular to a closed orbit in the energy shell). More
precisely, $A_k$ has the following expression~:
\begin{equation}
\label{amplitude}
  A_k=2(2\pi)^{3/2}\sqrt{\sin\theta_i\sin\theta_f}{\mathcal{Y}}_m(\theta_i)
  {\mathcal{Y}}_m(\theta_f)\frac 1{\sqrt{|m_{12}|}},
\end{equation}
where $\theta_{i,f}$ are respectively the initial and final angles (at
the nucleus) 
of the trajectory with the axis $z$ of the magnetic field.
The function ${\mathcal{Y}}_m(\theta)$ depends only on
the structure of the initial state and the polarization of the exciting light. 
Physically, it represents the angular distribution of the excited
electron leaving 
the nucleus. Explicit expressions for
excitation with $\pi$-polarized light from the $n=1$ (ground state) or 
$n=2$ are the following~\cite{Bogomolny89,Du88,Main97a}:
\begin{equation}
\label{ym}
  \begin{split}
    |\psi_i\rangle=|1s0\rangle & \qquad
    {\mathcal{Y}}_0(\theta)=-\pi^{-1/2}2^3e^{-2}\cos\theta \\
    |\psi_i\rangle=|2s0\rangle & \qquad
    {\mathcal{Y}}_0(\theta)=-(2\pi)^{-1/2}2^8e^{-4}\cos\theta \\
    |\psi_i\rangle=|2p0\rangle & \qquad 
    {\mathcal{Y}}_0(\theta)=(2\pi)^{-1/2}2^7e^{-4}(4\cos^2\theta-1).
  \end{split}
\end{equation}

The phase $\phi_k$ is given by~:
\begin{equation}
  \phi_k=S_k(\gamma,E)-\frac{\pi}2\mu_k+\frac{\pi}4,
\end{equation}
where $S_k$ is the action of the closed orbit. The index
$\mu_k$ takes into account the different phase shifts occurring
at conjugate points, at the nucleus and at the crossings 
of the $z$-axis. This index will be called Maslov index, even if it
is not the usual one.
More precisely, it is given by the 
following formula (for $m=0$)~\cite{Bogomolny89}:
\begin{equation}
  \label{eq:maslov}
  \mu=\nu_0+\nu_1+\nu_2+\nu_3,
\end{equation}
where $\nu_0$ is the number of conjugate points ($m_{12}=0$) along the 
closed orbit, $\nu_1$
is the number of points at which the velocity vanishes (only for self
retracing orbits), $\nu_2$ is the number of crossings of the $z$-axis
and $\nu_3$ is the number of times that an orbit leaves and reaches
the nucleus. The case of the orbit along the 
field (i.e. the $z$-axis) is special and requires a slight modification
(for details, see \cite{Main94,Bogomolny89}).  

The singularity in the classical equations of motion due to the
divergence of the Coulomb potential at $\mathbf{r=0}$ is regularized
using the semi-parabolic coordinates
$(u=\sqrt{r+z},\,v=\sqrt{r-z})$, giving rise to the following effective
classical Hamiltonian~\cite{Main97a,Englefield}:
\begin{equation}
  h=\frac 12 p_{u}^2+\frac 12 p_{v}^2-\epsilon(u^2+v^2)+\frac
  18u^2v^2(u^2+v^2)=2.
\end{equation}
One can show that, due to the potential shape, there is a deep analogy 
of the classical dynamics with the 4-disk
problem~\cite{Eckhard90,Hansen95,Tanner96} (i.e. the dynamics of a point
particle moving freely in the $(u,v)$ plane and bouncing
elastically off the walls  made by four disks at the corners of a square).
In the magnetized hydrogen atom, the role of the 4 disks are played
by the 4 potential hills between the valleys along the $u$ and $v$ axis.
Thus, the coding scheme of the 4-disk can be  
directly used~:
the $(u,v)$ plane is divided in four parts delimited by the
coordinates axis and each part is labeled with a number $\in\{1,2,3,4\}$ 
(see Fig.~\ref{orbit}). Then for each orbit, a sequence $s_1\dots s_n$ 
is built, corresponding to the sequence of bounces~\cite{bounce}
made by this orbit. At large positive scaled energy (above $\epsilon=0.33$),
it seems that this coding scheme works perfectly
well~\cite{Hansen95}. At lower scaled energy, 
some sequences of symbols become forbidden (pruning takes place); in the region
studied in this paper, the coding scheme is still quite efficient. It
allows us to derive few properties 
like the Maslov indices $\mu_k$ in a very simple
way~\cite{Eckhard90,Hansen95,Tanner96}.

Using scaling properties of the classical dynamics, the oscillating
part of the cross-section can be rewritten as~: 
\begin{equation}
  \label{scaledtrace}
  \begin{split}
  \sigma_{\mathrm{osc}}(\gamma,\epsilon)
  &=\frac{4\pi\omega}c\sum_k\gamma^{1/6}\sigma_k(\epsilon) \\
  &=\frac{4\pi\omega}c
  \sum_k\gamma^{1/6}\tilde{A}_k(\epsilon)
  \sin\left(\gamma^{-1/3}\tilde{S}_k(\epsilon)
    -\frac{\pi}2\mu_k(\epsilon)+\frac{\pi}4\right).
\end{split}
\end{equation}
The scaled
amplitudes $\tilde{A}_k$ are
functions of the scaled energy $\epsilon$ only (except for the orbit
along the field, which is varying like  $\gamma^{1/6}$). The exact expression
is directly derived from equation~\eqref{amplitude}~:
\begin{equation}
  \label{ampliang}
  \tilde{A}_k(\epsilon)=2(2\pi)^{3/2}\sqrt{\sin\theta_i\sin\theta_f}
  {\mathcal{Y}}_m(\theta_i)
  {\mathcal{Y}}_m(\theta_f)
  \frac 1{\sqrt{|\tilde{m}_{12}(\epsilon)|}}.
\end{equation}

Looking at
formula~\eqref{scaledtrace} one can see that, by fixing the scaled
energy $\epsilon$ and doing the Fourier transform of
$\gamma^{-1/6}\sigma(\gamma,\epsilon)$ 
with respect to the variable $\gamma^{-1/3}$, one must
obtain peaks at the scaled actions of the closed orbits~: this is the
well-known scaled spectroscopy \cite{Holle88}.

However, the preceding approach fails in case of a saddle-node
bifurcation at a given value $\epsilon_c$ of the scaled
energy. At the bifurcation, two orbits are created, but with $\tilde{m}_{12}$
coefficients equal to 0 and thus the scaled amplitude
$\tilde{A}_k$ of the two orbits are infinite. This divergence can be
regularized by a detailed study of the modifications induced in the
trace formula by the bifurcation~\cite{Kus93,Main97a}. More precisely, 
the contribution to the oscillating part of the cross-section due to the
orbits involved in the bifurcation becomes~:
\begin{multline}
  \label{airy}
  \gamma^{1/6}\sigma_{k}=2(2\pi)^{3/2}\sqrt{\sin\theta_i\sin\theta_f}
  {\mathcal{Y}}_m(\theta_i)
  {\mathcal{Y}}_m(\theta_f)
  \gamma^{1/9}\left(\frac{3\tilde{\sigma}}2\right)^{1/6}|\tilde{M}|^{-1/2}\\
  \times Ai\left((\frac{3\tilde{\sigma}}2)^{2/3}\gamma^{-2/9}
    (\epsilon_c-\epsilon)\right) 
    \sin
    \left(\gamma^{-1/3}(\tilde{S}(\epsilon_c))-\frac{\pi}2
    \mu^0\right),
\end{multline}
where $Ai(z)$ is the Airy function~\cite{Abramovitz}, $\tilde{\sigma}$
and $\tilde{M}$ are defined by local expansion of $\tilde{S}$ and
$\tilde{m_{12}}$ near the bifurcation, that is~:
\begin{equation}
\label{linearisation}
  \left\{\begin{array}{rll}
      \tilde{S}(\epsilon)_{\pm}&=\tilde{S}(\epsilon_c)
      &\pm\tilde{\sigma}
      \left(\epsilon-\epsilon_c\right)^{3/2} \\
      \tilde{m}_{12}(\epsilon)_{\pm}&=&\pm\tilde{M}
      \left(\epsilon-\epsilon_c\right)^{1/2}, \\
  \end{array}\right.
\end{equation}
where $\pm$ refers to the pair of orbits born above the bifurcation.
At fixed value of the scaled energy
$\epsilon>\epsilon_c$ and for very small values of
$\gamma$ which corresponds to the semi-classical limit, using 
the asymptotic behaviour of Airy function~\cite{Abramovitz}, one obtains~:
\begin{equation}
  \label{scab}
  \begin{split}
     \gamma^{1/6}\sigma_{k}&=\frac{4\pi\omega}c\gamma^{1/6}
     \tilde{A}_0(\epsilon)\sin 
    \left(\gamma^{-1/3}(\tilde{S}(\epsilon)_-)-\frac{\pi}2(\mu^0-
      \frac 12)
    \right) \\
    &+\frac{4\pi\omega}c\gamma^{1/6}\tilde{A}_0(\epsilon)\sin
    \left(\gamma^{-1/3}(\tilde{S}(\epsilon)_+)-\frac{\pi}2
    (\mu^0+\frac 12)\right)
  \end{split}
\end{equation}
with $\tilde{A}_0(\epsilon)$ is given by~:
\begin{equation}
\tilde{A}_0(\epsilon)=2(2\pi)^{3/2}\sqrt{\sin\theta_i\sin\theta_f}
  {\mathcal{Y}}_m(\theta_i)
  {\mathcal{Y}}_m(\theta_f)\frac 1{\sqrt{|\tilde{M}|
      |\epsilon-\epsilon_c|^{1/2}}}
\end{equation}
recovering thus the usual semi-classical contribution of the two
closed orbits created at the bifurcation.

Below the bifurcation and again using the 
asymptotic behaviour of Airy function~\cite{Abramovitz}, one obtains~:
\begin{equation}
  \label{scbb}
     \gamma^{1/6}\sigma_k = \frac{4\pi\omega}c\gamma^{1/6}\tilde{A}_0(\epsilon)
  \sin\left(\gamma^{-1/3}
  \tilde{S}(\epsilon_c)-\frac{\pi}2\mu^0\right)
  e^{-\gamma^{-1/3}\tilde{\sigma}(\epsilon_c-\epsilon)^{3/2}}.
\end{equation}
The contribution does not vanish, even if it is exponentially
decreasing when going away from the bifurcation or going to the semi-classical
limit $\gamma \rightarrow 0.$ 
It has exactly the
same functional dependence than for a usual closed orbit,  if one
allows the action to 
become complex $\tilde{S}(\epsilon) = \tilde{S}(\epsilon_c)-i \tilde{\sigma}
      \left(\epsilon_c-\epsilon\right)^{3/2},$ which is nothing
      but the continuation of Eq.~(\ref{linearisation}) across the bifurcation.
In fact, it can be interpreted as the contribution
of a ``ghost" closed orbit, living in a complexified phase
space~\cite{Kus93,Main97a}. 
 
\section{Harmonic inversion}
\label{harm}

This method is a powerful tool to extract Fourier components (phase
and amplitude) of a signal, with a better accuracy than one
obtained by the usual Fourier transform, which is limited by the total 
length of the signal. Especially, this method is very well
adapted to distinguish peaks closer than the Fourier resolution,
provided that they are well separated from the other ones. A complete
description of this method and some of its applications can be found in
references~\cite{Main97b,Mandelshtam97,Main98}.

Given a time signal $c(t)$, known in the interval $[0,T]$ and for
which we assume the following expression~:

\begin{equation}
  \label{eq:signal}
  c(t)=\sum_{n=1}^{N}a_n e^{-i\omega_n t},
\end{equation}

where $a_n$ and $\omega_n$ are the unknown amplitudes and frequencies to
be extracted from the signal. 

\begin{figure}
\centerline{\psfig{figure=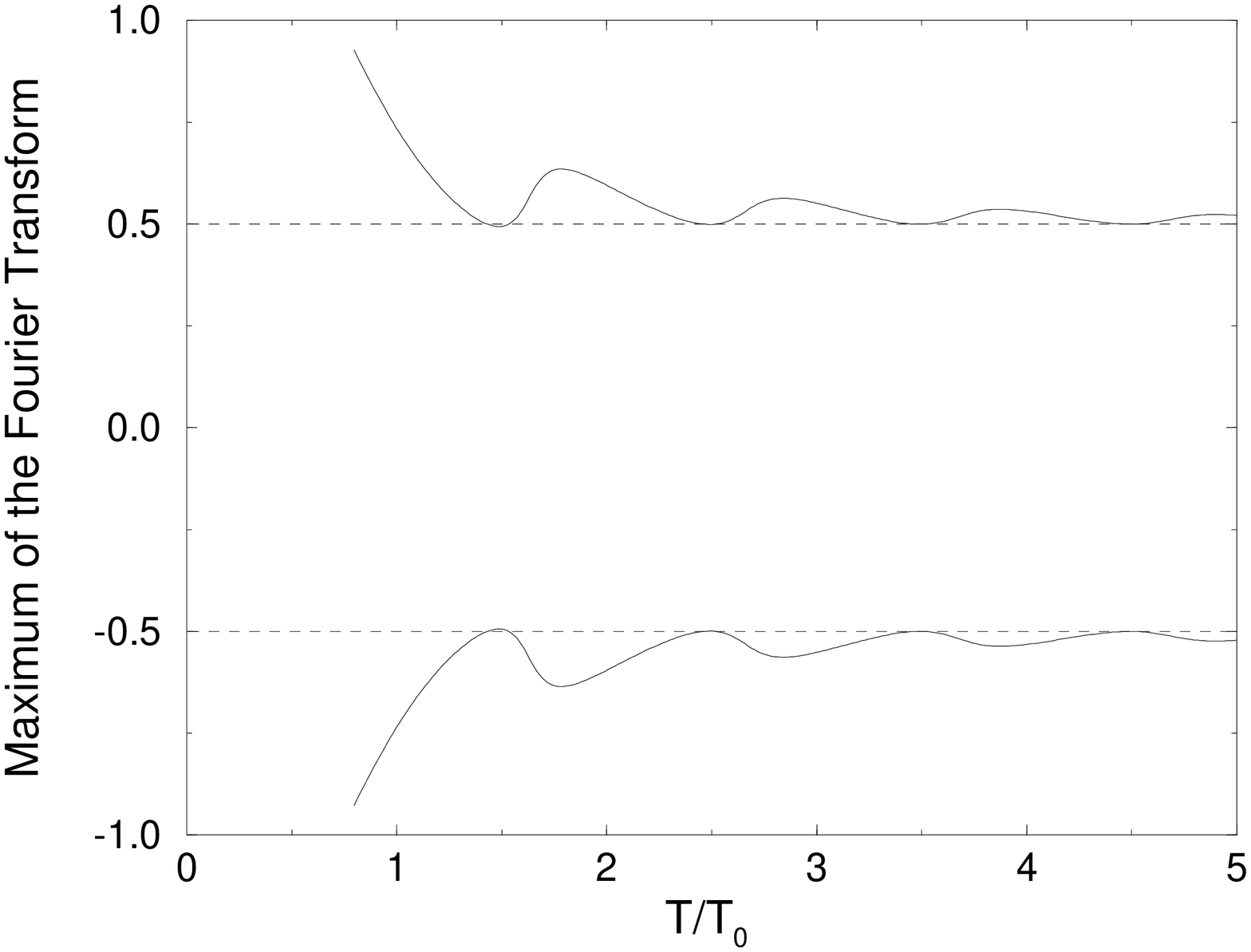,width=15cm,angle=-90}}
\bigskip
\caption{\label{twopeaks} Effective frequencies extracted from the
  signal $c(t)=\exp{(i\delta\omega t/2)}+\exp{(-i\delta\omega t/2)}$ using 
  finite Fourier transform, i.e. $\int_0^Tdt\,c(t)\exp{i\tilde{\omega}t}$,
  as functions of $T$. For large values of $T$ (much larger than
  $T_0=2\pi/\delta\omega$) we recover the right values of
  $\tilde{\omega}$, that is $\pm \delta\omega/2$ (dashed lines). On
  the contrary, for  
  $T$ of the order of $T_0$, the frequencies given by the Fourier
  transform are substantially shifted from the correct values. For
  example, for $T=2T_0$, the distance between the two peaks is
  enlarged by 20 percent. }
\end{figure}

Of course, these can be obtained using standard Fourier transform, but 
then the resolution is limited by the interval length $T$, adding an
artificial width equal to $2\pi/T$ to each Fourier peak. Especially,
for frequencies closer one to each other than $4\pi/T$, their values
does not correspond to the respective maxima in the Fourier spectrum,
as one can see from the following simple example~:
\begin{equation}
  \label{eq:deltaw}
\begin{split}
  c(t)&=ae^{-i(\omega_0+\frac 12\delta\omega)t}+
  be^{-i(\omega_0-\frac 12\delta\omega)t} \\
  &=ae^{-i\omega_+t}+be^{-i\omega_-t}
\end{split}
\end{equation}
where $a$ and $b$ are real numbers. The modulus square of the
time limited Fourier transform (i.e. 
$f(\omega)=\int_0^T\!\!dt\,c(t)e^{i\omega t}$), is given by~:
\begin{multline}
  \label{eq:modtf}
  |f(\omega)|^2=a^2T^2\sinc^2{(\omega-\omega_+)\frac T2}+
  b^2T^2\sinc^2{(\omega-\omega_-)\frac T2} \\
   + 2abT^2\cos{(\omega_+-\omega_-)\frac T2}
    \sinc{(\omega-\omega_+)\frac T2}\sinc{(\omega-\omega_-)\frac T2}
\end{multline}
where the function $\sinc(x)$ is the usual $\sin(x)/x$. As expected,
it is only a function of $\tilde{\omega}=\omega-\omega_0$~:
\begin{multline}
  \label{modtft}
  |f(\tilde{\omega})|^2=a^2T^2\sinc^2{(\tilde{\omega}-\frac
    12\delta\omega)\frac T2}+b^2T^2\sinc^2{(\tilde{\omega}+\frac
    12\delta\omega)\frac T2} \\
  +2abT^2\cos{\delta\omega\frac T2}
  \sinc{(\tilde{\omega}-\frac 12\delta\omega)\frac T2}
  \sinc{(\tilde{\omega}+\frac 12\delta\omega)\frac T2}
\end{multline}
For 
$T$ large enough (i.e. much larger than $T_0=2\pi/\delta\omega$), the two
peaks of the Fourier transform are well separated and we are able to
extract from this Fourier transform the right values of the two
frequencies $\tilde{\omega}_+=\delta\omega/2$ and
$\tilde{\omega}_-=-\delta\omega/2$. On the contrary, as one can see  
in Fig.~\ref{twopeaks}, when $T$ is of the order of
$2\pi/\delta\omega$, the effective positions of the peaks are less and
less well 
defined. Even for $T=4\pi/\delta\omega$ (i.e. theoretical Fourier
resolution two times as small as $\delta\omega$), the distance between the two
peaks is enlarged by 20 percent. Furthermore, how this shift is distributed 
between the two peaks depends strongly of the relative amplitude
$a/b$. For example if $b$ is 5 times as small as $a$, then $b$ is
shifted by 85 percent of the total shift, whereas $a$ is only shifted
by 15 percent. 

It becomes worse and worse as the number of frequencies in the signal
increases,  
which shows that the usual Fourier transform is of little interest to
extract accurate information from a limited signal. On the contrary,
the harmonic inversion can bypass this Fourier limitation.

The main idea of the harmonic inversion is to construct an abstract evolution
operator $U(t)$ and an abstract initial state $|\Phi_0\rangle$ such that $c(t)$
is given by $\langle\Phi_0|U(t)|\Phi_0\rangle$, where
$\langle\quad.\quad\rangle$ is a complex symmetric inner product (i.e.
$\langle\psi|$ is just the transpose of $|\psi\rangle$, without
complex conjugation). The eigenvalues of
this operator are the $e^{-i\omega_nt}$, showing that $U$ is not an
unitary operator ($\omega_n$ being generally complex).
Then by diagonalizing  its
matrix representation in a suitable basis, it is possible to extract
with a much higher accuracy than the Fourier limit these eigenvalues; 
the associated eigenvectors are then used to find the amplitudes
$a_n$. One possible basis is a Krylov basis ~: $|n\rangle=U(n\delta
t)|\Phi_0\rangle$,  where $\delta t$ is a short time interval and $n$ is
an integer ($0\leq n\leq N$, $N$ defining the basis size). Thus, the matrix
representation is simply given by $A_{nn'}=\langle
n|U(\delta t)|n'\rangle=c\bigl((n+n'+1) \delta t\bigr)$. The basis vectors
being not orthogonal 
($B_{nn'}=\langle n|n'\rangle=c\bigl((n+n') \delta t\bigr)$),
one has to solve the symmetric generalized eigenvalues problem~:
\begin{equation}
  \label{gep}
  A|\phi\rangle=e^{-i\omega\delta t}B|\phi\rangle
\end{equation}

However, this simplest basis choice is not satisfactory for efficient
numerical purpose, because all coefficients in the matrices have the 
same order of magnitude, such that a eigenvector will have significant 
overlap with all basis states. Furthermore, all frequencies $\omega_n$ 
are equally represented, which means that $N$ has to be of the order
of the total number of frequencies, which can be very large.
For this reason, another basis, using
the filtering properties of the Fourier transform is introduced.
The choice of the basis is done by selecting an interval 
$[\omega_{\mathrm{inf}},\omega_{\mathrm{sup}}]$ in which one wants to
extract frequencies. For frequencies $\omega_j$ in this interval, one builds
a new basis $|\psi_j\rangle=\sum_n e^{in\omega_j\delta t}|n\rangle$. From
the signal $c(t)$, only frequencies in the interval will give
significant contributions to the basis vectors. Thus, one can take a
much smaller basis, of the order of the number of frequencies in this
interval. Moreover, the matrix structure of the evolution operator in
this basis in essentially banded, that is matrix coefficients are
decreasing rapidly when going away from the diagonal. The expressions
of the coefficients and other details about the numerical resolution
of the new  generalized eigenvalues problem can be found
in~\cite{Main97b,Mandelshtam97,Main98}.

\begin{figure}
\centerline{\psfig{figure=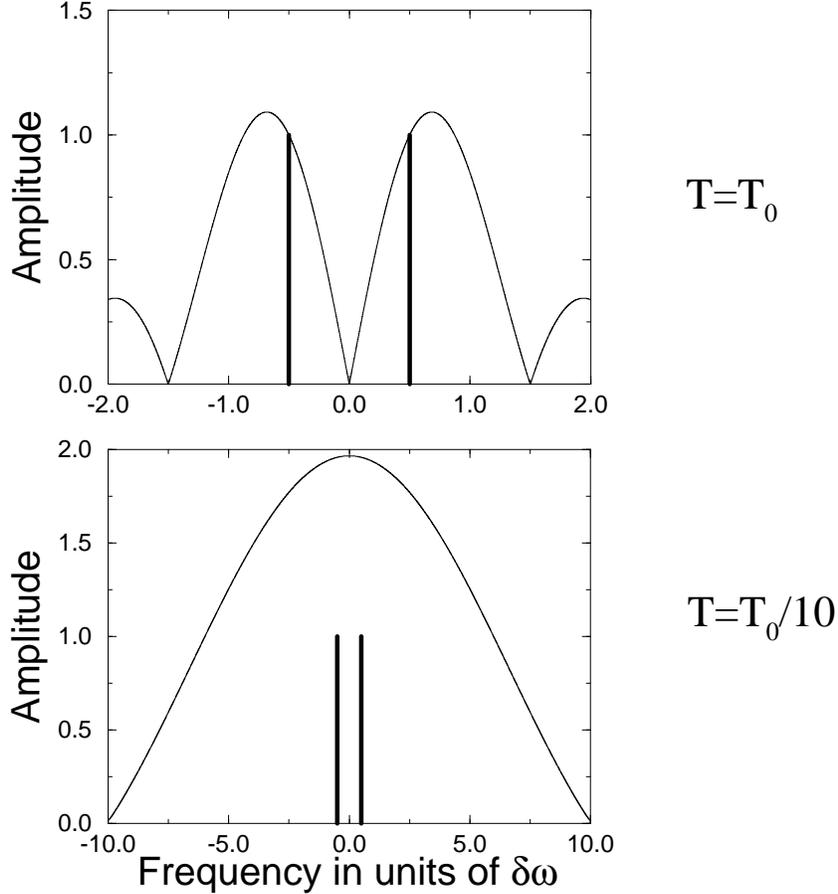,width=15cm,angle=-90}}
\vspace{2cm}
\caption{\label{invsim} Comparison between the usual Fourier transform and 
  the harmonic inversion technique for the signal 
  $c(t)=\exp{(i\delta\omega t/2)}+\exp{(-i\delta\omega t/2)}$. For both
  graphs, the continuous line is  
  the usual Fourier transform $|F(T)|/T$ (T is the length of the
  available signal $c(t)$), whereas the vertical lines
  are at the frequencies given by the harmonic inversion, their
  heights corresponding to the associated amplitude (also obtained by
  harmonic inversion). The upper graph is made for
  $T=T_0=2\pi/\delta\omega$, which means that the Fourier resolution is 
  of the order of the peak separation. Again, we recover the fact that
  the effective positions given by the Fourier transform are quite
  shifted from the exact values and the amplitudes are also not
  correct. On the contrary, all values (amplitudes and positions)
given by the harmonic inversion are in a perfect agreement. The lower
graph is made for $T=T_0/10$, i.e. the Fourier resolution is 10 times
as large as the peak separation, so that the Fourier transform is
almost constant in the interval $[-1,1]$ and thus gives no information 
on the position of the peaks. In contrast, the harmonic
inversion method is still working, with a very good accuracy (better
than $10^{-7}$).}
\end{figure}

A first test of this method is to apply it to the preceding
example. The results are shown in Fig.~\ref{invsim}. For both
graphs, the continuous line is modulus of the usual Fourier transform
divided by $T$ and the vertical segments
are the amplitudes given by the harmonic inversion. Sizes of the Krylov basis
from $2$ to $50$ have been used, although $2$ would have been
enough, to emphasize that even when using an over-complete basis (which 
would be the standard case, as the number of frequencies is unknown) the 
results given by the harmonic inversion are not affected.
The upper graph has been made for a signal
length equal to $T_0=2\pi/\delta\omega$, i.e. the theoretical Fourier
resolution is the distance between the two peaks. The bottom graph is
made for a signal length $T_0/10$. From both pictures, it appears
clearly that the harmonic inversion is much more efficient than the
usual Fourier transform~: for $T=T_0$, the Fourier transform still
shows two peaks, but shifted by more than 20 percent from the exact
positions and for $T=T_0/10$, the two peaks have collapsed in a single
peak, whose width is much larger than $\delta\omega.$
 
On the contrary, values extracted by the harmonic
inversion are very well localized, at the right frequencies. More
precisely, the accuracy on different parameters (position and
amplitude) are shown in table~\ref{invsimtab}. Of course, this example 
is rather simple, but even for signal length such that the Fourier
resolution would be ten times as large as the peak separation and for
relative amplitudes differing by two orders of magnitude, the
accuracies are better than $10^{-6}$ on the frequencies and $10^{-5}$
on the amplitudes. For more complicated situations, the accuracy is
expected to be worse than in the present example, but it will be still 
much better than what we would be able to extract from the Fourier transform.

\section{Results}
\label{res}

How one can apply the harmonic inversion in the case of the hydrogen atom in
a magnetic field is obvious if one compares
equations~\eqref{eq:signal} and \eqref{scaledtrace}~: $\gamma^{-1/3}$
corresponds to time $t$, the scaled action $\tilde{S}_k(\epsilon)$ to 
(minus) $\omega_n$ and 
$\tilde{A}_k\exp\left(-i\frac{\pi}2\mu_k+i\frac{\pi}{4}\right)$ to $a_n$. 
For negative 
scaled energy $\epsilon$, the function
$\tilde{\sigma}(\gamma,\epsilon)$ 
is a just a sum of Dirac delta functions~:
\begin{equation}
  \label{eq:dirac}
  \tilde{\sigma}(\gamma,\epsilon)=\frac{4\pi\omega}c\sum_{n=0}^{\infty}
  \tilde{f}_n(\epsilon)
  \delta(\gamma^{-1/3}-\gamma_n^{-1/3})
\end{equation}
where $\gamma_n^{-1/3}$ are the eigenvalues of the scaled quantum Hamiltonian 
(i.e. $\epsilon$ being fixed, energy $E$ quantization, because of
equation~\eqref{eet}, is equivalent to $\gamma^{-1/3}$
quantization). $\tilde{f}_n$ are the associated (squared) excitation
matrix elements. 

That we explicitly know the functional dependency of
$c(t=\gamma^{-1/3})$ can be used to simplify the implementation of the 
harmonic inversion. Indeed, the latter was written in the case of a
general signal obtained either numerically or experimentally and known only
at equally spaced values of time. On the contrary, in the 
present case, we are able to compute $c(t)$ for any time $t$. A
straightforward method would be to compute $c(t)$ only on a grid. But
because of $\delta$ functions we would have to add an artificial width
to each peak  for numerical purpose, as it is done in
ref~\cite{Main98}. The other way round is to modify a little bit the
generalized eigenvalues problem~\eqref{gep}, in which the matrix
representation of the evolution operator $U(t)$ was used,
to obtain directly a generalized eigen-equation for the operator
$\Omega$,  defined by $U(t)=e^{-i\Omega t}$, which is very close to
what was done in the original articles of
Neuhauser~\cite{Neuhauser90,Wall95}. If one writes the eigensystem
in the following form~:
\begin{equation}
  \label{mgep}
  \tilde{A}|\phi\rangle=-i\omega\tilde{B}|\phi\rangle
\end{equation}
then the coefficients of $\tilde{A}$ are given by~:
\begin{equation}
\begin{split}
  \label{coeffA}
  \tilde{A}(\phi_1,\phi_2) &= \\
  (\phi_1\ne\phi_2)\qquad&\phantom{=}\frac{1}{\phi_1-\phi_2}\Bigl[\phi_2
  \int_0^{T/2}\!\!dt\,c(t)e^{it\phi_2}-
  \phi_1\int_0^{T/2}\!\!dt\,c(t)e^{it\phi_1} \\
  &-\phi_2e^{i\frac T2(\phi_1-\phi_2)}\int_{T/2}^T\!\!dt\,c(t)e^{it\phi_2}
  +\phi_1e^{i\frac T2(\phi_2-\phi_1)}\int_{T/2}^T\!\!dt\,c(t)e^{it\phi_1}
  \Bigr] \\
  (\phi_1=\phi_2)\qquad &-\Bigl[\int_0^{T/2}\!\!dt\,c(t)e^{it\phi_1}
  +i\phi_1\int_0^{T/2}\!\!dt\,c(t)te^{it\phi_1}
  -\int_{T/2}^T\!\!dt\,c(t)e^{it\phi_1} \\
  &+iT\phi_1\int_{T/2}^T\!\!c(t)e^{it\phi_1}
  -i\phi_1\int_{T/2}^T\!\!dt\,c(t)te^{it\phi_1} \Bigr]
\end{split}
\end{equation}
where $\phi_1$ and $\phi_2$ are two frequencies taken in the interval 
$[\omega_{\textrm{inf}},\omega_{\textrm{sup}}]$ in which one wants to
extract the $\omega_n$. The matrix elements of $\tilde{B}$ have
similar expressions~:
\begin{equation}
  \label{coeffB}
  \begin{split}
   \tilde{B}(\phi_1,\phi_2) &= \\
   (\phi_1\ne\phi_2)\qquad&\phantom{=}\frac{i}{\phi_1-\phi_2}\Bigl[
   \int_0^{T/2}\!\!dt\,c(t)e^{it\phi_2}-\int_0^{T/2}\!\!dt\,c(t)e^{it\phi_1}\\
   &-e^{i\frac{T}{2}(\phi_1-\phi_2)}\int_{T/2}^T\!\!dt\,c(t)e^{it\phi_2}
   +e^{i\frac{T}{2}(\phi_2-\phi_1)}\int_{T/2}^T\!\!dt\,c(t)
   e^{it\phi_1}\Bigr] \\
   (\phi_1=\phi_2)\qquad&\phantom{=}\int_0^T\!\!dt\,c(t)e^{it\phi_1}
   (\frac{T}2-|\frac{T}2-t|)
   \end{split}
\end{equation}
Insertion of signal~\eqref{eq:dirac} in the preceding expressions is
straightforward and leads us to replace all integrals with sums on the
available eigenenergies $\gamma^{-1/3}_n$.

Using the preceding implementation of the harmonic inversion, we will
focus on the so-called X1 bifurcation \cite{Holle88}, 
which is of the saddle-node type~: at the 
bifurcation  energy $\epsilon_c=-0.11544216$, two closed orbits are
created, one stable and one instable. The important point is that
these two orbits have classical actions very well separated from all
other closed orbits, so that their evolution with the scaled energy
can be easily followed (both on the experimental and theoretical
points of view).

All quantum properties (energy levels and 
excitation matrix elements) have been obtained by numerical diagonalization
of sparse matrices, which are representations of the full Hamiltonian
in sturmian basis (for further details see ref.~\cite{Delande}). Thus, 
for each value of the scaled energy $\epsilon$, we have $\approx 10000$
eigenvalues, giving a length signal equal to $\approx 120$.
$\epsilon$ values are ranging from $-0.15$ to $-0.07$ with $0.001$
step. For each
scaled energy below $\epsilon_c$, the complex scaled action
and the amplitude of the ghost peak have been extracted, whereas for
scaled energy above $\epsilon_c$, actions of both real orbits created
at the bifurcation and their amplitudes have been extracted. 

\begin{figure}
\centerline{\psfig{figure=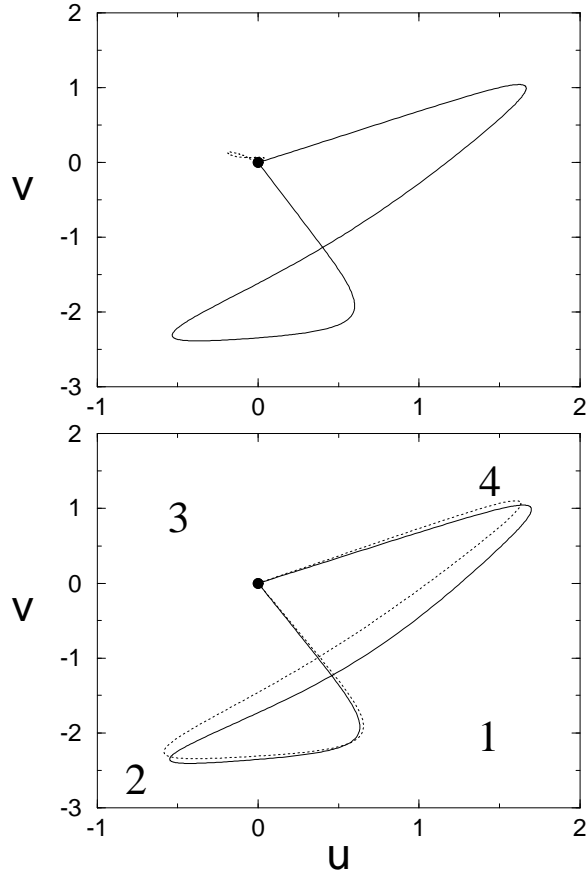,width=15cm,angle=-90}}
\vspace{1cm}
\caption{\label{orbit} Closed trajectories of the electron of an hydrogen atom
in a magnetic field in the  $(u,v)$ (semi-parabolic coordinates)
configuration space.  On the top is shown the
  complex ghost orbit associated with the X1 peak 
  experimentally observed in the scaled spectroscopy of the hydrogen atom in
  a magnetic field at scaled energy $\epsilon=-0.14$ 
  \protect\cite{Holle88}. The continuous
  line is the real part of the trajectory, whereas the dotted line is
  the imaginary part. Eventually, this complex trajectory collapses
  with its complex conjugate in a saddle-node bifurcation at
  $\epsilon_c=-0.11544216$, from which two real orbits are
  created, shown on the bottom at scaled energy $\epsilon=-0.11$. The 
  dashed line corresponds to the $S_-$ orbit created at the
  bifurcation, whereas the continuous line corresponds to the $S_+$
  orbit. From this plot, one can easily deduce the code of each
  orbit, that is the sequence of bounces in the different quarters of
  space (1,2,3,4) defined by the coordinate axis~: for $S_-$, the code
  is $124$ and $1214$ for $S_+$ (one additional
  bounce on the $(u>0,v<0)$ part of the plane).}  

\end{figure}

To emphasize the agreement with the semi-classical
formula~\eqref{scab}~\eqref{scbb}, we have 
also computed, for each value of the scaled energy, the classical quantities
(scaled action, Lyapunov exponent and Maslov indices). For scaled
energy below $\epsilon_c$, the ghost orbits are found by extending the
classical equations of motion in complex plane, which also makes time to be 
complex. More precisely, for real closed orbits, one has to determine
two parameters (i.e. the initial direction of the orbit and its time
length), whereas for complex closed orbits one has to determine four
parameters (i.e. the complex initial direction and the  complex final
time). However, integration for complex time requires to integrate
along paths in the 
complex plane; it is far from obvious that the
(complex) classical quantities are functions of the final time
only. For simple systems, like a one-dimensional double well potential, 
one can show that the classical quantities are meromorphic functions 
of the final time, but to our knowledge, there is no exact proof 
in the case of the hydrogen atom in a magnetic field. 
Still,  the potential being a
polynomial function of the coordinates $(u,v)$, the structure of the
equations of motion is such that the classical quantities are locally
analytic, so that small deviations from a given path lead to the same
results. Thus, even if a complete study of all singularities and branch
points in the complex plane would be needed to justify this approach,
one restricts integration of classical equations of motion on straight 
lines in the complex time plane starting from
the origin, the integration time being simply the length along the line. In 
that case, the usual Runge-Kutta method can still be used. 
For example,
the ghost orbit associated with the X1 bifurcation is shown in
Fig.~\ref{orbit}(top). The scaled energy is $\epsilon=-0.14$. The
continuous line is the real part, that is in 
the plane $\bigl({\mathrm{Re}}(u),{\mathrm{Re}}(v)\bigr)$, whereas the
dotted line is the imaginary 
part (i.e. in the plane
$\bigl({\mathrm{Im}}(u),{\mathrm{Im}}(v)\bigr)$. On the bottom are
shown the two real orbits (at scaled energy $\epsilon=-0.11$) created
at the bifurcation. The dashed line corresponds to the $S_-$ orbit and 
the continuous line to the $S_+$ orbit. The fact that the only
difference between the two trajectories is the additional bounce
(i.e. another conjugate point) in
the $(u>0,v<0)$ part of the space for the longest orbit ($S_+$)
explains that the Maslov indices  associated with these orbits are
$\mu_-=\mu_0$ and $\mu_+=\mu_0+1$. If we use the coding scheme of the
4-disk problem discussed above, the two orbits involved in the bifurcation
have respectively
codes 124 and 1214 (see Fig.~\ref{orbit}), from which (using
Eq.~\eqref{eq:maslov}) we calculate
the Maslov indices $\mu_0=8$ and $\mu_0+1=9.$

\begin{figure}
\centerline{\psfig{figure=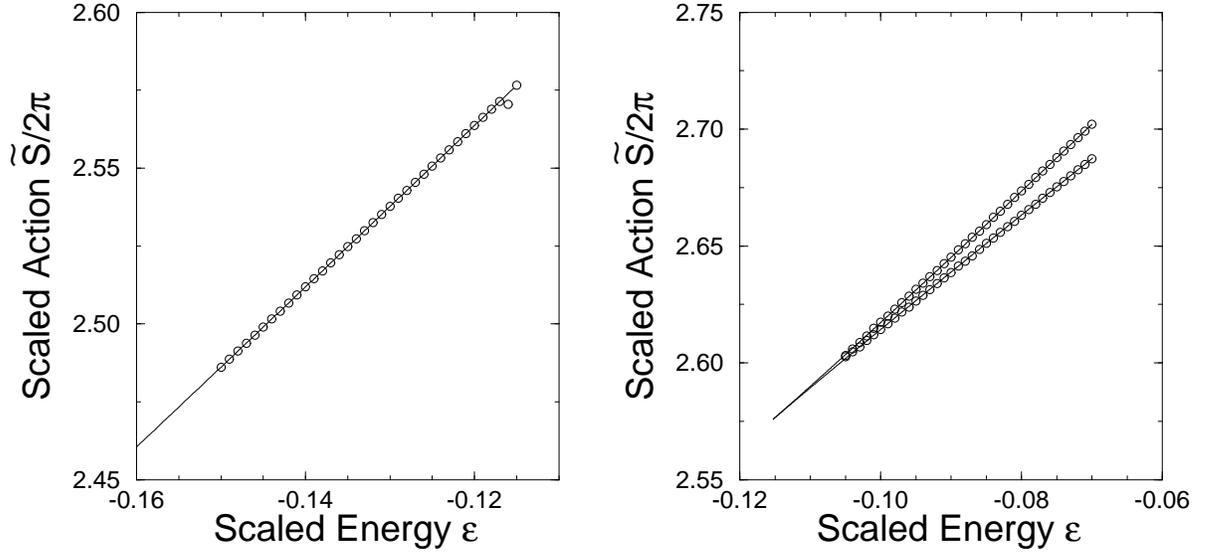,width=15cm,angle=-90}}
\bigskip
\caption{\label{action} Real part of the scaled actions as function of 
  the scaled energy $\epsilon$. The continuous lines are the 
  classical results, whereas circles are extracted from the quantum
  dynamics using harmonic inversion. One can follow the two real orbits 
  born at the saddle-node
  bifurcation, even for separation much smaller than the standard Fourier
  limitation ($\approx 0.01$).}
\end{figure}

\begin{figure}[htbp]
\centerline{\psfig{figure=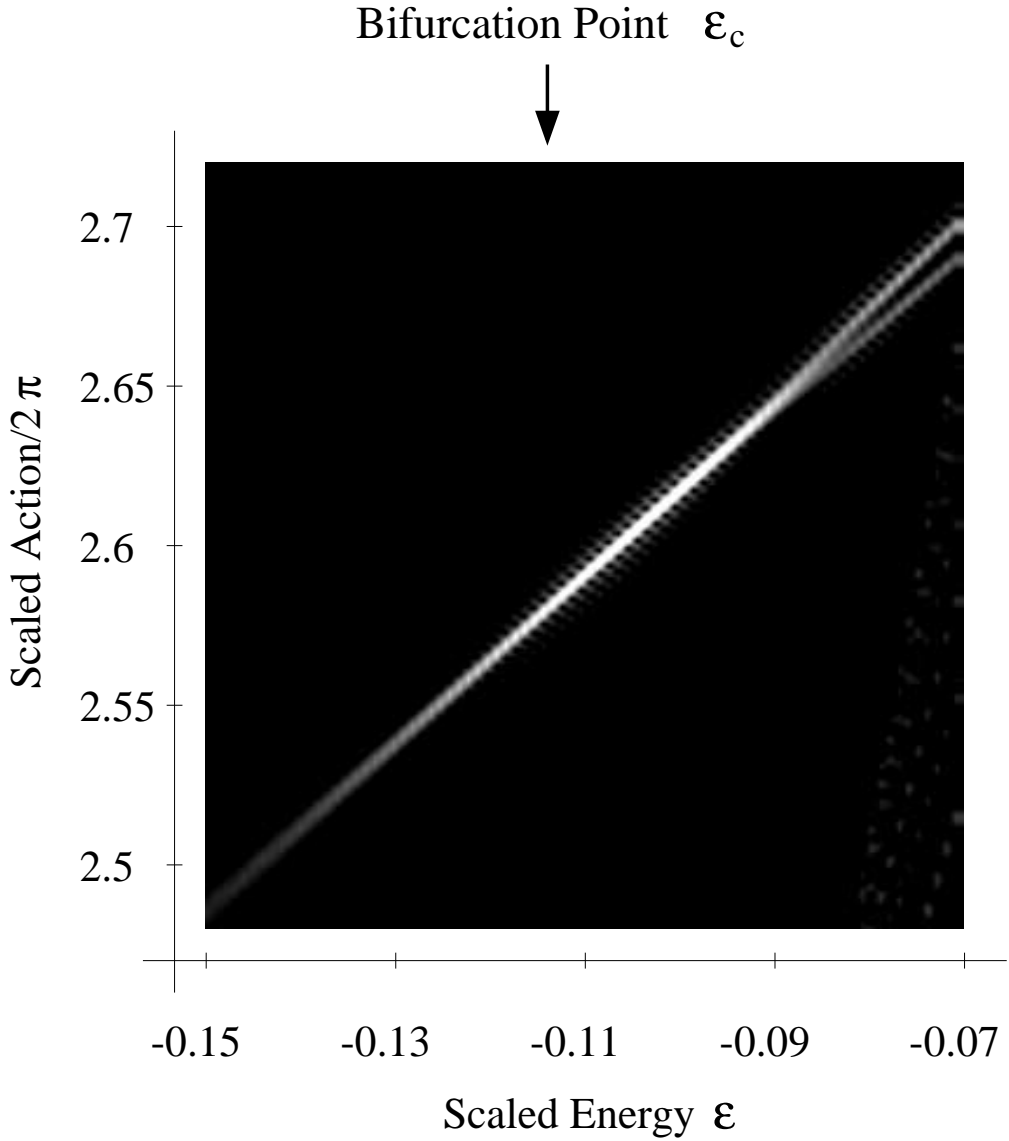,height=15cm}}
\bigskip
\caption{\label{ftx1} Usual scaled Fourier transform 
  (see equation~\eqref{uft}) of the
  photo-excitation cross-section of the hydrogen atom in a magnetic field,
   in the
  $(\epsilon,\tilde{S}/2\pi)$-plane.  Below the bifurcation
  $\epsilon_c=-0.11544216,$ one clearly sees the contribution of the
  ghost orbit.  
  However, it would be very hard to extract the relevant 
  information about the complex orbit, namely the imaginary part of
  the action, from the width of the peak. Especially, it should vanish
  like $(\epsilon_c-\epsilon)^{3/2}$ near the bifurcation, which is
  not the case in the figure. Actually, it is dominated by the
  broadening due to the limited range of the Fourier transform. 
  Above the bifurcation, the
  actions of the two
  real orbits created at the bifurcation are separated by the same
  power law. But the separation cannot be observed on this plot
  below $\epsilon\approx-0.085$, which is already too far from the
  bifurcation point to observe this power law.} 
\end{figure}

\begin{figure}
\centerline{\psfig{figure=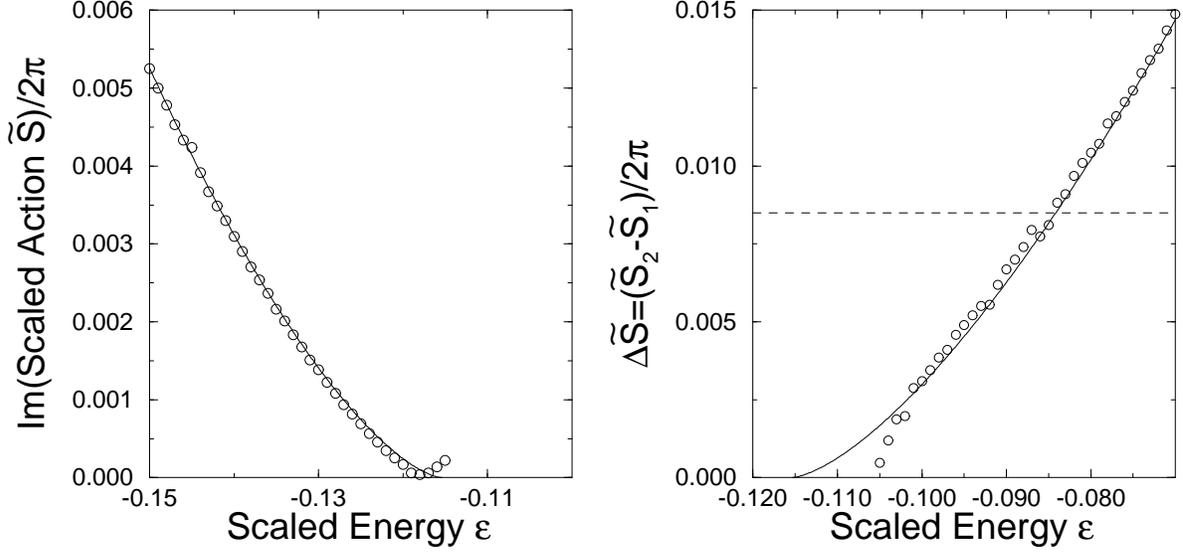,width=15cm,angle=-90}}
\bigskip
\caption{\label{deltaS} On the left is the imaginary part of the scaled
  action of the ghost
  orbit (below the bifurcation) and on the right is the separation in action
  between the two real orbits above the bifurcation. For both graphs,
  we include levels up to
  $\gamma^{-1/3}_{\mathrm{max}}=120$, so that the theoretical Fourier
  limitation would be $\approx 8.5\times 10^{-3}$, but 
  (see text),  
  the actual resolution is at least twice as large. On the contrary,
  as it appears clearly on both figures, the limitations of the
  harmonic inversion are much smaller, which allows us to emphasize
  the good agreement between the semi-classical predictions and the
  exact quantum results.}
\end{figure}

The first comparison is shown by Fig.~\ref{action} where scaled
actions of the orbits are shown as functions of the scaled
energy. Below the bifurcation (on the left), is shown the real part of 
the action and above the bifurcation (on the right), are shown 
the actions of the two orbits born at the bifurcation. The continuous
lines are the classical calculation, whereas the circle are the values
extracted from the quantum data. The agreement is excellent, even very 
close to the bifurcation point. This
agreement is emphasized in Fig.~\ref{deltaS}, where on the left side, is
shown the imaginary part of action of the ghost orbit, and, on the
right side, is 
shown the difference $\Delta S=(\tilde{S}_2-\tilde{S}_1)/2\pi$. Here
again, the agreement is good, especially if one compares with
the Fourier resolution which is $\approx0.0085$. It shows clearly that 
it would have been impossible to extract the right behaviour of neither
the width of the ghost peak nor the separation in the actions of the
real orbits  
where it is the most interesting, that is as close as possible 
of the bifurcation. This clearly appears in Fig~\ref{ftx1} where the usual
Fourier transform, that is
\begin{equation}
\label{uft}
|F(\epsilon,\tilde{S})|=\left|\int_0^{\gamma^{-1/3}_{\mathrm{max}}}\!\!
d(\gamma^{-1/3})\,\gamma^{-1/6}\sigma(\gamma,\epsilon)
e^{-i\tilde{S}\gamma^{-1/3}}\right|
\end{equation}
is drawn in the $(\epsilon,\tilde{S}/2\pi)$-plane.
It is obvious that one can not
distinguish the two orbits for scaled energy $\epsilon$ closer to the
bifurcation than $\epsilon=-0.085$. 
Furthermore, even if for scaled energy greater than
$-0.085,$ the two real orbits seem to be far enough one from each
other to be seen on the Fourier transform, we know from the simple
example shown in section~\ref{harm} that even for theoretical Fourier 
resolution twice better than the peaks separation, the extracted
positions are still shifted by $\approx 20$ percent. This tells us that 
quantitative comparison with the classical dynamics would have been
possible only for scaled energy larger than $-0.07$. But at that
point, other peaks corresponding to a family of orbits start to
overlap with these two 
peaks, so that it would be hardly impossible to extract any
quantitative information. Also, it would be difficult, at present time, 
to have  a Fourier resolution twice better only by increasing the signal
length~:  as the density of states is proportional to $\gamma^{-1/3}$,
the number of states below a given $\gamma^{-1/3}$ is proportional to
$\gamma^{-2/3}$, and thus to obtain a signal length twice as long, one 
would have to compute more than 40000 levels, which is at 
the edge of the present numerical possibilities (i.e. it would require 
32 times as much CPU time and 8 times as much memory).

\begin{figure}[htbp]
\centerline{\psfig{figure=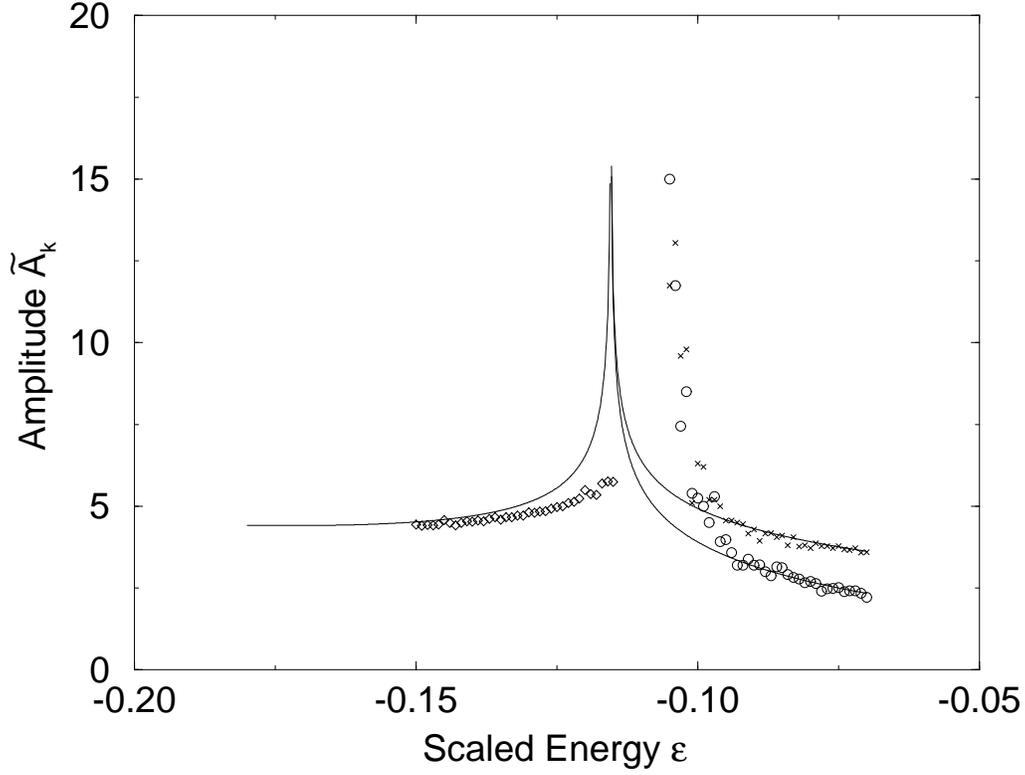,width=15cm,angle=-90}}
\vspace{1cm}
\caption{\label{amplit} Comparison of quantum amplitudes
  $\tilde{A}_k(\epsilon)$ (see equations.~\eqref{scaledtrace}~\eqref{ampliang})
  extracted using the  
  harmonic inversion, with the semi-classical
  prediction. The continuous lines are the
  amplitudes calculated from the classical quantities (stability,
  initial and final angle). Diamond (ghost orbit), crosses and circles 
  (real orbits) are the amplitudes extracted from the exact quantum results.
  On both sides of the bifurcation, two regimes
  appear. For scaled energy away enough from the
  bifurcation ($|\epsilon-\epsilon_c|>0.02$) the semi-classical
  predictions and the quantum results agree well. On the contrary
  for scaled energy too close to the bifurcation point, the
  deviation is large.}
\end{figure}

After having compared the frequencies, one can compare the amplitudes
with the semi-classical predictions. The results are shown in
Fig.~\ref{amplit}. The continuous lines are  the
individual semi-classical amplitudes as obtained from 
equation~\eqref{ampliang}, with the ${\mathcal{Y}}_m$ function corresponding 
to transition from the $|2s0\rangle$ initial state, with
$\pi$-polarized light (see equation~\eqref{ym}).
On both sides of the bifurcation point $\epsilon_c$, two
different regimes 
clearly appear. For $|\epsilon-\epsilon_c|>0.02$, the
agreement with the 
semi-classical prediction is quite good, even if not as good as
for the actions. On the contrary for energies too close to the
bifurcation point (i.e. $|\epsilon-\epsilon_c|<0.02$), the
discrepancy is large, 
showing thus that the semi-classical approximation is no more valid. 
This can
be easily understood from 
equation~\eqref{airy}, which shows that the semi-classical limit
(i.e. $\gamma^{-1/3}\rightarrow +\infty$) is in competition with the
limit $\epsilon\rightarrow\epsilon_c$. In other words,
for fixed 
value of $\gamma^{-1/3}$, one goes out of the semi-classical limit
when approaching the bifurcation. The break point corresponds to the
argument of the Airy function of the order of one, which for
$\gamma^{-1/3}=120$ (the largest available value) gives
$\epsilon-\epsilon_c\approx 0.01$, in agreement with the
results. For $\epsilon>\epsilon_c$, the argument of
the Airy function being directly related to the separation in action 
$\Delta\tilde{S}$ (see equation~\eqref{linearisation}), the
preceding argument corresponds exactly to the fact that, for fixed
scaled energy value, the phase shift
$2\pi\gamma^{-1/3}\Delta\tilde{S}$  
has to be large enough ($\gtrsim 2\pi$), so that the two orbits become
distinguishable from the quantum point of view.
A very simple test of the fact that we cannot use any asymptotic
expansion of the Airy function, is to consider  the amplitude
exactly at the bifurcation point. In that case, the Airy function is
only taken at the value 0, for any value of $\gamma^{-1/3}$. More
precisely, the term
$B_{k_c}\sin\left(\gamma^{-1/3}\tilde{S}(\epsilon_c)-\frac{\pi}2\mu^0\right)$
in the oscillating part of the cross-section (see
equation~\eqref{airy}) becomes~: 
\begin{multline}
  \label{eq:airy(0)}
  \gamma^{1/6}\sigma_k = \frac{4\pi\omega}{c} \gamma^{1/9}\ 2(2\pi)^{3/2}
\sqrt{\sin\theta_i\sin\theta_f}
  {\mathcal{Y}}_m(\theta_i)
  {\mathcal{Y}}_m(\theta_f)
  \left(\frac{3\tilde{\sigma}}2\right)^{1/6} \\
  \times |\tilde{M}|^{-1/2}
  Ai(0)\sin\left(\gamma^{-1/3}\tilde{S}(\epsilon_c)-\frac{\pi}2\mu^0\right)
\end{multline}

From this equation, one can see that it is possible to extract the
amplitude from the quantum data using harmonic inversion, provided
that the signal is multiplied by $\gamma^{-1/9}$. For this purpose,
$7000$ levels have been computed at scaled energy
$\epsilon=\epsilon_c$. The amplitude extracted from this
data is $2.94$, in a very good agreement with the theoretical value $2.951$
given by equation~\eqref{eq:airy(0)}.

\begin{figure}[htbp]
  \centerline{\psfig{figure=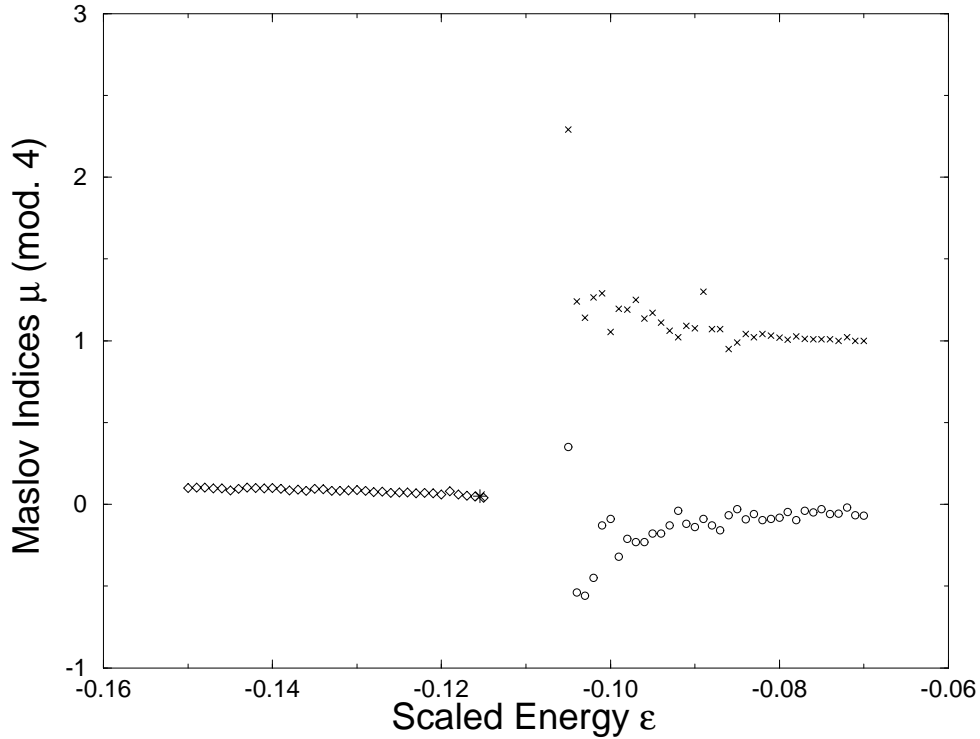,width=15cm,angle=-90}}
\bigskip
\caption{\label{maslov} Maslov indices (see
    equation~\eqref{eq:maslov} in text)
    associated with each orbit extracted from the exact quantum
    results. Above the bifurcation point $\epsilon_c$, we recover the
    semi-classical prediction that Maslov indices of the two real
    orbits created in a saddle-node bifurcation are $\mu_0$ and
    $\mu_0+1$, where $\mu_0$ is also the Maslov index of the ghost 
    orbit. Furthermore, this value $\mu_0=0$ (modulo 4) agrees with the one
    obtained from Eq.~(\eqref{eq:maslov}), giving $\mu_0=8$. The
    symbol (+) correspond to the $\mu_0$ value extracted exactly at the
    bifurcation point, using a specific harmonic inversion technique
    (see text), also in perfect agreement with the theoretical
    predictions.} 
\end{figure}

Having compared the modulus of the amplitude, we now turn to its
phase, i.e. Maslov indices.  As the Maslov indices occur only in the phases
with a multiplicative factor $\pi/2,$ they can only be measured modulo 4.
Fig.~\ref{maslov} shows the nice
agreement between the Maslov indices extracted from harmonic inversion
of the quantum signal and the semi-classical predictions. Indeed, from
equation~\eqref{scab}, the Maslov indices of the two real orbits 
($\epsilon>\epsilon_c$) are
respectively $\mu_0$ and $\mu_0+1$ (with $\mu_0=8,$ i.e. 0 modulo 4) 
whereas it is $\mu_0$ for the
ghost orbit ($\epsilon<\epsilon_c$)
which is exactly what is depicted
by the figure. Exactly at the bifurcation point, with the modified
harmonic inversion described in the previous paragraph, we obtain a 
Maslov index of $0.05$, again
in perfect agreement with the theoretical value $\mu_0=0$ (modulo 4).

Even if the agreement between the different quantities is excellent,
it is not as exceptional as in the simple
example~\eqref{eq:deltaw}. The possible explanation comes from the
fact that one assumes the expression for the signal to be a sum of
exponentials with constant coefficients. This is clearly broken in the 
case of the semi-classical approximation because of the smooth
contribution in either the density of state or the oscillator strength 
(i.e. Thomas-Fermi like terms) and also because of the remaining 
terms in the asymptotic expansion in $\hbar$. All this will contribute 
to (slowly) varying coefficients, which will make the harmonic
inversion method less efficient. This is clearly emphasized by the
behaviour of the amplitudes for scaled energy too close to the bifurcation
for which one cannot use anymore the asymptotic expansion of the  Airy
function.

\section{Conclusions}

In summary, we have shown that,  using the harmonic inversion method,
the properties (actions, Maslov indices and 
stabilities) of the classical orbits involved in the photo-ionization
cross-section of the hydrogen atom in a magnetic field can be extracted with
a much better accuracy than the usual Fourier transform, and this even for 
scaled energies close to a bifurcation point. Below the
bifurcation, the contribution of a ghost orbit has been emphasized by
showing that the behaviour of the imaginary part of its action is in
perfect agreement with the classical predictions. For scaled energy
above the bifurcation, we have been able to distinguish the contributions
of the two orbits created at the bifurcation and we have also shown
the perfect agreement with the semi-classical prediction for the frequencies,
amplitudes and phases of the modulations. We have
also emphasized the non-semiclassical behaviour of the amplitudes 
too close to the bifurcation point. 

Laboratoire Kastler Brossel is laboratoire de l'Universit{\'e} Pierre et Marie
Curie et de l'Ecole Normale Sup{\'e}rieure, unit{\'e} mixte de
recherche 8552 du CNRS. CPU time on a Cray C90 computer has been provided
by IDRIS.

\begin{table}
\caption{\label{invsimtab} Accuracy on parameters extracted from 
$c(t)=a_0e^{-i(\omega_0+\delta\omega/2)t}+b_0e^{-i(\omega_0-\delta\omega/2)t}$ 
using harmonic inversion, for signal length  $T=10T_0$, $T=T_0$ and
$T=T_0/10$, where $T_0=2\pi/\delta\omega$. This table emphasizes the
excellent accuracy of the method. For $T>T_0$, it  is of the
order of the machine precision.  For signal length ten times
as short as the theoretical Fourier resolution, the accuracy is still 
of 
the order of $10^{-7}$ on the frequency positions and $10^{-6}$ on the 
amplitudes. Even for very peculiar situations (one peak 100 times as
small as the other one and signal length 10 times as short as the
theoretical Fourier resolution), the accuracy is still very good.}

\begin{tabular}{ccccc}
$a_0=b_0$ & $|\omega_+-\omega_+^0|/\omega_+^0$ & 
$|\omega_--\omega_-^0|/\omega_-^0$ & $|a-a_0|/a_0$ & $|b-b_0|/b_0$ \\
\tableline
$T=10T_0$ & $<10^{-14}$ & $<10^{-14}$ & $<10^{-14}$ & $<10^{-14}$ \\
$T=T_0$ & $\lesssim 10^{-13}$ & $\lesssim 10^{-13}$ & $\lesssim 10^{-12}$
& $\lesssim 10^{-12}$ \\
$T=T_0/10$ & $\lesssim 10^{-7}$ & $\lesssim 10^{-7}$ & 
$\lesssim 10^{-6}$ & $\lesssim 10^{-6}$ \\ \tableline
$100a_0=b_0$ &&&& \\ \tableline
$T=10T_0$ & $<10^{-14}$ & $<10^{-14}$ & $<10^{-14}$ & $<10^{-14}$ \\
$T=T_0$ & $\lesssim 10^{-12}$ & $\lesssim 10^{-14}$ & $\lesssim 10^{-11}$
& $\lesssim 10^{-13}$ \\
$T=T_0/10$ & $\lesssim 10^{-6}$ & $\lesssim 10^{-8}$ & 
$\lesssim 10^{-5}$ & $\lesssim 10^{-7}$ \\
\end{tabular}
\end{table}


\begin{thebibliography}{99}
\bibitem{bounce} For the magnetized hydrogen atom, these bounces are
``soft" bounces on the potential hills, which, as shown in
\protect{\cite{Hansen95}}, 
can be associated with zeros of the matrix element $m_{12}$ of the 
monodromy matrix.
\end{thebibliography}
\end{document}